\documentclass[12pt]{article}

\newcommand{\beq}{\begin{eqnarray}}
\newcommand{\eeq}{\end{eqnarray}}

\begin{document}

\begin{center}
\hspace{10cm}

\begin{flushright}
BCCUNY-HEP 2004/01\\
February 2, 2004\\
Revised, February 24,2004
\end{flushright}

\vspace{15pt}
{\large \bf
GAUGE-INVARIANT COORDINATES ON GAUGE-THEORY ORBIT SPACE}

\vspace{30pt}

{\bf Peter Orland}$^{\rm a.}$$^{\rm b.}$\footnote{Work supported in
part by a PSC-CUNY 
Research Award}

\vspace{8pt}
 
\begin{flushleft}
a. Physics Program, The Graduate School and University Center,
The City University of New York, 365 Fifth Avenue,
New York, NY 10016, U.S.A.
\end{flushleft}

\begin{flushleft}
b. Department of Natural Sciences, Baruch College, The 
City University of New York, 17 Lexington Avenue, New 
York, NY 10010, U.S.A., orland@gursey.baruch.cuny.edu 
\end{flushleft}

\vspace{40pt}

{\bf Abstract}
\end{center}

\noindent 
A gauge-invariant field is found which describes physical 
configurations, i.e.
gauge orbits, of non-Abelian gauge theories. This is 
accomplished with non-Abelian generalizations of 
the Poincar\'{e}-Hodge decomposition formula for one-forms. In
a particular sense, the new field is dual to the
gauge field. Using this field as a coordinate, the
metric and intrinsic curvature are discussed for  
Yang-Mills orbit space for the (2+1)- and (3+1)-dimensional
cases. The sectional, Ricci and scalar curvatures are all formally 
non-negative. An expression for the
new field in terms of the Yang-Mills connection is
found in 2+1 dimensions. The measure on Schr\"{o}dinger 
wave functionals is found in both 2+1 and
3+1 dimensions; in the former case, it resembles 
Karabali, Kim and Nair's measure. We briefly discuss
the form of the Hamiltonian in terms of the dual field and
comment on how this is relevant to the mass
gap for both the (2+1)- and (3+1)-dimensional cases.

\vspace{30pt}

\section{Introduction}
\setcounter{equation}{0}
\renewcommand{\theequation}{1.\arabic{equation}}

Many field theorists have speculated as to whether QCD can
be solved by the substitution of the Yang-Mills connection
by gauge-invariant degrees of freedom. The source
of this speculation is that the space of physical degrees
of freedom or gauge orbits $\cal M$ is the quotient
of connection space $I\!\!\Lambda \!\!\!\!\!-$ by 
gauge transformations $\cal G$,
neatly written as
\begin{eqnarray}
{\cal M}=I\!\!\Lambda\!\!\!\!\!\!- /{\cal G}\;.
\nonumber
\end{eqnarray}
A gauge orbit
O is an equivalence class containing all gauge transforms
of  some gauge field $A_{k}^{g}=g^{-1}A_{k}g^{-1}+{\rm i}g^{-1}\partial_{k}g$. Schr\"{o}dinger
wave functionals depend on orbits - not gauge fields. Reformulating 
Yang-Mills theory in terms of gauge orbits can
be accomplished
in principle (except on a set of measure 
zero) by gauge fixing and eliminating
Gribov copies 
\cite{gribov,singer,bab-via2,orl1996,kud-mor-orl,orl-sem}. The quest for a gauge-invariant
formalism of gauge theories began decades ago \cite{lots-and-lots}. A 
radical idea is to find a string-model
reformulation; remarkably, such a reformulation 
been quite successful for non-asymptotically-free
gauge theories \cite{maldacena}.

In this paper, we find a relatively simple way of 
characterizing coordinate charts of $\cal M$
as hypersurfaces in {\it I}$\!\Lambda \!\!\!\!\!-$. These hypersurfaces
are cross sections of the fiber bundle, and there are Gribov
copies. Many of the geometric properties of orbit space
are much easier to understand, however, than 
with ordinary gauge-fixing
procedures. For example, the metric tensor and intrinsic curvature
are simple to obtain. A single-valued
definition of the coordinates exists, at least in
2+1 dimensions, which solves the Gribov 
problem. In 2+1 
dimensions, there appears to be a connection with the remarkable methods 
of Karabali and Nair \cite{kar-nair} and of Karabali, Kim and 
Nair \cite{kar-kim-nair}. 

We work only with SU($N$) gauge fields in Hamiltonian
formalism in $D$ space and one time dimension. The gauge connection
$D_{j}=\partial_{j}-{\rm i}A_{j}$ where $j=1,\dots,D$ and $A(x)_{j}
=\sum_{\alpha} A(x)_{j}^{\alpha} t_{\alpha}$ is
a traceless, Hermitian $N\times N$ matrix field. The basis
of the Lie algebra consists of generators $t^{\alpha}$, with
$[t_{\alpha}, t_{\beta}]=
i\sum_{\gamma}f_{\alpha \beta}^{\gamma} t_{\gamma}$. For 
any
$N\times N$ matrix field $Q$, we define ${\cal D}_{j}Q=
[D_{j},Q]$ and the gauge field in the adjoint representation
by ${\cal A}_{j}$ by ${\cal D}_{j}
=\partial_{j} -{\rm i}{\cal A}_{j}$. The field strengths in the
fundamental representation and the adjoint representation
are $F_{jk}={\rm i}[D_{j},D_{k}]$, and 
${\cal F}_{jk}={\rm i}[{\cal D}_{j},{\cal D}_{k}]$, 
respectively. 

We adopt a notation suited to the discussion
of components of 
vectors and tensors on connection space and orbit space. We
will write the gauge field as
$$A^{(x,j,\alpha)}=A(x)_{j}^{\alpha}\;,$$
and Lie-Algebra-valued scalar fields similarly, e.g.
$$\Phi^{(x,\alpha)}=\Phi(x)^{\alpha}\;.$$

Let us define {\it I}$\!\Lambda \!\!\!\!\!-$ to consist of square-integrable connections
on infinite space (not space-time) ${{\rm I}\!{\rm R}}^{\rm D}$, with
square-integrable field strengths $F$. 

The metric on orbit space $\cal M$, is
\begin{eqnarray}
d\rho^{2}=\int 
G_{(x,j,\alpha)(y,k,\beta)}\;
\delta A^{(x,j,\alpha)}
\delta A^{(y,k,\beta)}\;,
\label{first-metric}
\end{eqnarray}
where the integration denotes a sum over all repeated
indices (including $x$ and $y$!) and the metric is
\begin{eqnarray}
G_{(x,j,\alpha)(y,k,\beta)}=\left[\delta_{jk}\delta^{\alpha \beta}
-{\cal D}_{j}\left( P \frac{1}{{\cal D}^{2}}\right) {\cal D}_{k}
\right]\delta^{\rm D}(x-y)\;, \label{first-metric-tensor}
\end{eqnarray}
where $P$ denotes the principle value. This metric is the projection
which removes
those variations of the gauge field 
of the form 
\begin{eqnarray}
\delta A^{(x,j,\alpha)}=
{\cal D}^{\alpha}_{\;\;\;\beta}\Omega^{ (\!(x,\beta)\!) }\;.
\label{infinitesimal-gauge-trans}
\end{eqnarray}
The reason for the double-bracket notation $(\!(x,\beta)\!)$ is to distinguish variations which
are gauge transformations from those which are not.

The metric (\ref{first-metric}) was argued by Babelon and Viallet \cite{bab-via2} to be 
the infinitesimal version
of the distance function between an orbit ${\rm O}$ containing 
gauge field $A$ and an orbit ${\rm O}^{\prime}$ containing gauge field
$A^{\prime}$
\begin{eqnarray}
\rho({\rm O},{\rm O}^{\prime} )^{2}=\frac{1}{2}\inf_{g\in {\cal G}} \int d^{\rm D}x \;
\{[A^{g}(x)]^{2}-A^{\prime}(x)^{2}\}\;. \nonumber
%\label{largemetric}
\end{eqnarray}
This was proven for
appropriately defined gauge fields as Hilbert-space vectors in reference \cite{orl1996}. There it
was also shown by a simple argument that the size of orbit space is unbounded in 3+1 dimensions - the
potential energy $\int {\rm Tr }F_{ij}^{2}$ can even be arbitrarily small at sufficiently large distances
from a pure gauge. This indicates that 
proofs that orbit space is bounded \cite{zwanziger} do not apply to the thermodynamic limit. Feynman
attempted to explain 
confinement and the mass gap in 2+1 dimensions by approximating this distance on
configurations of small magnetic energy \cite{feyn}. The object of Feynman's investigation was
to show that such configurations lie within a finite diameter. Orland
and Semenoff showed that Feynman's distance 
approximations were incorrect, but were actually able to calculate 
some distances exactly \cite{orl-sem}, supporting this idea.

The virtue of the expression on the right-hand
side of (\ref{first-metric-tensor}), namely that is has 
gauge transformations as zero modes
(\ref{infinitesimal-gauge-trans}), is also its curse. Metric tensors with
zero eigenvalues 
are rather difficult
to work with. A formalism to deal with the geometry from this viewpoint has been discussed in
reference \cite{orl1996}. An
alternative, which we believe to be more useful, is presented below.

In the next section we present a non-Abelian decomposition formula for differential one-forms. This provides
a powerful new gauge-fixing method in Section 3, and we use this to study the gauge-fixed
hypersurface in the space of connections. Gribov singularities are realized as the appearance of
harmonic one-forms in the decomposition formula. We elaborate on some issues concerning gauge invariance
in Section 4. In Section 5, we find formal expressions for the extrinsic and (intrinsic)
curvature tensor of the hypersurface. The Ricci and scalar curvatures are discussed for arbitrary
space dimension in Section 6. We briefly examine the Yang-Mills Hamiltonian formulated on the
hypersurface in Section 7. In Section 8, we show how to explicitly obtain the hypersurface coordinate
in terms of the gauge connection in 2+1 dimensions; this is used to find a Hamiltonian formalism which
strongly resembles that of Karabali, Kim and Nair \cite{kar-nair, kar-kim-nair} in Section 9. We introduce
a dual Hamiltonian and very briefly discuss how the
Yang-Mills mass gap arises in Section 10. Our conclusions and directions for 
further work are summarized in
Section 11.

\section{Non-Abelian Poincar\'{e}-Hodge decomposition formulas}
\setcounter{equation}{0}
\renewcommand{\theequation}{2.\arabic{equation}}

Let us briefly recall the Poincar\'{e}-Hodge decomposition
formula. Any differential form $C$ may be decomposed as
\begin{eqnarray}
C=d\Phi+\ast [d (\ast \Omega)] +h\;, \label{p-h}
\end{eqnarray}
where $h$ is a harmonic form; by which it is meant
that $h$ is both closed $dh=0$, and co-closed $d \ast h=0$. For
two dimensions of space, and a 
a one-form $C=\sum_{j}C^{(x,j)}\cdot dx^{j}$ (\ref{p-h}) may be written as 
\begin{eqnarray}
C^{(x,j)}=\epsilon^{jk}\partial_{k} \Phi^{(x)}+ 
\partial_{j} \Omega^{(x)}
+h^{(x,j)}\;,  \nonumber
%\label{abelian-2d}
\end{eqnarray}
and for three space dimensions
\begin{eqnarray}
C^{(x,j)}=\epsilon^{jkl}\partial_{k} \Phi^{(x,l)}+ 
\partial_{j} \Omega^{(x)}
+h^{(x,j)}\;. \nonumber
%\label{abelian-3d}
\end{eqnarray}
Harmonic forms are particularly important in finite volumes. The dimensions of the space
of harmonic forms are
the Betti numbers of the manifold. For some special finite-volume manifolds, namely
star-shaped regions, there are no
harmonic forms, hence from (\ref{p-h}) 
any closed form $C$, $dC=0$ must be exact $C=d\Phi$.

What is especially significant about the decomposition of forms is that it is a decomposition
into subspaces of Hilbert space. Consider a second one-form in two space dimensions:
\begin{eqnarray}
C^{\prime\; (x,j)}=\epsilon^{jk}\partial_{k} \Phi^{\prime\;(x)}+ 
\partial_{j} \Omega^{\prime \;(x)}
+h^{\prime\;(x,j)}\;.  \nonumber
\end{eqnarray}
Then the inner product between $C$ and $C^{\prime}$ decomposes into three pieces:
\begin{eqnarray}
\left< C^{\prime} \vert C\right>= \int C^{\prime\;(x,j)}C^{(x,j)}
=\int \Phi^{\prime \;(x)} (-\partial^{2}) \Phi^{(x)}
+\int \Omega^{\prime \;(x)} (-\partial^{2}) \Omega^{(x)}
+\int h^{\prime \;(x,j)}h^{(x,j)}\;. \nonumber
\end{eqnarray}
A similar expression holds for the inner product in three space dimensions.

Consider a non-Abelian infinitesimal functional variation
$\delta A^{(x,j,\alpha)}$ which is covariantly
divergenceless, i.e. ${\cal D}^{\gamma}_{\;\;\;\alpha}
\delta A^{(x,j,\alpha)}=0$. For our purposes, we
must also assume that the Laplacian
on scalars ${\cal D}^{2}$ has no zero modes. We shall discuss this issue
further at the end of Section 4. We
say that $\delta A$ is non-Abelian exact in two space dimensions if
\begin{eqnarray}
\delta A^{(x,j,\alpha)}= \int L^{(x,j,\alpha)}_{\;\;\;\;\;\;(y,\beta)}
\;\delta \Phi^{(y,\beta)}\;, \label{2d-ym-exact}
\end{eqnarray}
where we have introduced the linear mapping
\begin{eqnarray}
L^{(x,j,\alpha)}_{\;\;\;\;\;\;(y,\beta)}
=\frac{\delta A^{(x,j,\alpha)}}{\delta \Phi^{(y,\beta)}}
=\left[ \epsilon^{jk}{\cal D}_{k}+{\rm i}
{\cal D}_{j} \frac{1}{{\cal D}^{2}} {\cal F}_{12}
\right]^{\alpha}_{\;\;\beta} \delta^{2}(x-y)\;,
\label{2d-ym-exterior-deriv}
\end{eqnarray}
and the integration is over $y$, with summation over $\beta$.
The expressions
${\rm i}{\cal D}_{j}$ and $\frac{1}{{\cal D}^{2}}$ 
in (\ref{2d-ym-exterior-deriv})
are to be understood
as self-adjoint
operators; they act on everything to the right. We do not
discuss the domain of $L$, but are
confident that it can be 
determined \cite{foot1}. We say
that $\delta A$ is non-Abelian exact in three space dimensions if
\begin{eqnarray}
\delta A^{(x,j,\alpha)}= L^{(x,j,\alpha)}_{\;\;\;\;\;\;(y,l,\beta)}
\;\delta \Phi^{(y,l,\beta)}\;, \label{3d-ym-exact}
\end{eqnarray}

\noindent
where
\begin{eqnarray}
L^{(x,j,\alpha)}_{\;\;\;\;\;\;(y,l,\beta)}
=\frac{\delta A^{(x,j,\alpha)}}{\delta \Phi^{(y,l,\beta)}}
=\left[ \epsilon^{jkl}{\cal D}_{k}+{\rm i}
{\cal D}_{j} \frac{1}{{\cal D}^{2}} (\ast {\cal F})^{(x,l)}
\right]^{\alpha}_{\;\;\beta} \delta^{3}(x-y)\;,
\label{3d-ym-exterior-deriv}
\end{eqnarray}
and the components of the magnetic field are
$(\ast {\cal F})^{(x,l)}=\frac{1}{2}\epsilon^{l m n} {\cal F}_{mn}$,
in the standard notation of Hodge. 

A non-Abelian exact variation ${\delta A}$ in either two or three space dimensions satisfies
${\cal D}\cdot \delta A=0$. This can be readily checked for (\ref{2d-ym-exact})
and (\ref{3d-ym-exact}).

Before going on to the
non-Abelian generalization of the decomposition
formula, let us take stock of the progress we have made. The
expressions
(\ref{2d-ym-exact}),
(\ref{2d-ym-exterior-deriv}), (\ref{3d-ym-exact})
and (\ref{3d-ym-exterior-deriv}) are powerful
tools for understanding the geometry
of orbit space. We can write the metric (\ref{first-metric}) (\ref{first-metric-tensor}) 
for two-space-dimensional
orbits
as
\begin{eqnarray}
d\rho^{2}=\int 
G_{(x,\alpha)(y,\beta)}\;
\delta \Phi^{(x,\alpha)}
\delta \Phi^{(y,\beta)}\;, \label{2d-metric}
\end{eqnarray}
where
\begin{eqnarray}
G_{(x,\alpha)(y,\beta)}=\int L^{(z,j,\gamma)}_{\;\;\;\;\;\;(x,\alpha)}
L^{(z,j,\gamma)}_{\;\;\;\;\;\;(y,\beta)}
= 
\left(-{\cal D}^{2}+ 
{\cal F}_{12}\frac{1}{-{\cal D}^{2}} {\cal F}_{12}
\right)_{\alpha \beta}\delta^{2}(x-y)\;, 
\label{2d-metric-tensor}
\end{eqnarray}
which is a non-negative quadratic form. 

From (\ref{2d-metric}) we see that $\delta \Phi^{(x,\alpha)}$
may be interpreted as the functional variation of a scalar
field $\Phi^{(x,\alpha)}$. It is this scalar field which
is a natural gauge-invariant set of coordinates.

In three dimensions, there is a subtlety, namely that the operator
$L^{(x,j,\alpha)}_{\;\;\;\;\;\;(y,l,\beta)}$ has zero 
modes. The quantity $\delta \Phi^{(x,l,\alpha)}$
may be interpreted as the functional variation of a vector
field $\Phi^{(x,l,\alpha)}$. Physically, this field is a new gauge field
which is {\em dual} to the
original gauge field. The zero modes of 
$L^{(x,j,\alpha)}_{\;\;\;\;\;\;(y,l,\beta)}$ are the dual
gauge transformations. These gauge transformations are 
complicated and will not be discussed in detail here. It is
no clear whether any harm results to the physics by fixing
or even
breaking the dual gauge invariance. For the time being we fix the dual gauge invariance by excluding
$l=3$ from $L^{(x,j,\alpha)}_{\;\;\;\;\;\;(y,l,\beta)}$. This
is a ``dual axial-gauge condition" in which $\Phi^{(x,3,\alpha)}$
is set to a constant. The dual-gauge-fixed vector field 
$\Phi^{(x,l,\beta)}$ is a coordinate we can use on orbit space
in three dimensions.

The metric in three space dimensions is
\begin{eqnarray}
d\rho^{2}=\int 
G_{(x,l,\alpha)(y,m,\beta)}\;
\delta \Phi^{(x,l, \alpha)}
\delta \Phi^{(y,m,\beta)}\;, \label{3d-metric}
\end{eqnarray}
where $l, m=1,2$ and
\begin{eqnarray}
G_{(x,l,\alpha)(y,m,\beta)}=
\int L^{(z,j,\gamma)}_{\;\;\;\;\;\;(x,l,\alpha)}
L^{(z,j,\gamma)}_{\;\;\;\;\;\;(y,m,\beta)} 
\;\;\;\;\;\;\;\;\;\;\;\;\;\;\;\;\;\;\;\;\;\;\;\;\;\;\;\;\;
\;\;\;\;\;\;\;\;\;\;\;\;\;\;\;\;\;\;\;\;
\nonumber 
\end{eqnarray}
\begin{eqnarray}
=\left[-{\cal D}^{2}\delta^{lm} +
{\cal D}_{m}{\cal D}_{l} 
+ (\ast {\cal F})_{l}\frac{1}{-{\cal D}^{2}}(\ast {\cal F})_{m}
\right]_{\alpha \beta}\delta^{3}(x-y)\;, \nonumber
\end{eqnarray}
\begin{eqnarray}
=\left[-{\cal D}^{2}\delta^{lm} +
{\cal D}_{l}{\cal D}_{m} -{\rm i} {\cal F}_{lm}
+ (\ast {\cal F})_{l}\frac{1}{-{\cal D}^{2}}(\ast {\cal F})_{m}
\right]_{\alpha \beta}\delta^{3}(x-y)\;, \nonumber
\label{3d-metric-tensor}
\end{eqnarray}
which is a non-negative symmetric quadratic form. 

The non-Abelian decomposition formula in two space dimensions
is
\begin{eqnarray}
\delta A^{(x,j,\alpha)}=
\int L^{(x,j,\alpha)}_{\;\;\;\;\;\;(y,\beta)} \delta \Phi^{(y,\beta)}
+\int{\cal D}^{(x,j,\alpha)}_{\;\;\;\;\;\;(\!(z,\sigma)\!)}
\delta \Omega^{(\!(z,\sigma)\!)} + h^{(x,j,\alpha)}\;,
\label{2d-ym-p-h}
\end{eqnarray}

\noindent
where

\begin{eqnarray}
{\cal D}^{(x,j,\alpha)}_{\;\;\;\;\;\;(\!(z,\sigma)\!)}
=({\cal D}_{j})^{\alpha}_{\;\;\;\sigma}\delta^{2}(x-z) \;, \nonumber
\end{eqnarray}
and
where the vector $h^{(x,j,\alpha)}$
satisfies the non-Abelian ``harmonicity" conditions, which we call
non-Abelian closed and non-Abelian co-closed, respectively:
\begin{eqnarray}
\int ({\cal D}^{\dagger})^{ (\!(z,\sigma)\!) }_{\;\;\;\;\;\; (x,j,\alpha) }\;
h^{(x,j,\alpha)}=0 \;, \label{2d-ym-closed}
\end{eqnarray}
\begin{eqnarray}
\int (L^{\dagger})^{(y,\beta)  }_{\;\;\;\;\;\; (x,j,\alpha) }\; 
h^{(x,j,\alpha)}=0 \;. \label{2d-ym-co-closed}
\end{eqnarray}
Notice that $\delta \Omega^{(\!(z,\sigma)\!)}$ is a gauge transformation. 

If 
there is another variation $\delta A^{(x,j,\alpha)}$, of $A^{(x,j,\alpha)}$, we may decompose it
as
\begin{eqnarray}
\delta A^{\prime\;(x,j,\alpha)}=
\int L^{(x,j,\alpha)}_{\;\;\;\;\;\;(y,\beta)} \delta \Phi^{\prime \;(y,\beta)}
+\int{\cal D}^{(x,j,\alpha)}_{\;\;\;\;\;\;(\!(z,\sigma)\!)}
\delta \Omega^{\prime \;(\!(z,\sigma)\!)} + h^{\prime\;(x,j,\alpha)}\;,
\nonumber
\end{eqnarray}

\begin{flushleft}
which yields the decomposition of the inner product:
\end{flushleft}
\begin{eqnarray}
\int \delta A^{\prime\;(x,j,\alpha)}\delta A^{(x,j,\alpha)}=
\int G_{(u, \alpha)(v,\beta)} \delta \Phi^{\prime \;(u,\alpha)}
\delta \Phi^{(v,\beta)}
\!&\!+\!&\!  \int (-{\cal D}^{2})_{(\!(z,\gamma )\!)(\!(w,\rho )\!)}
\delta \Omega^{\prime \;(\!(z,\gamma)\!)} 
\delta \Omega^{(\!(w,\rho)\!)} \nonumber \\
\!&\!+\!&\!\int h^{\prime\;(x,j,\alpha)}h^{(x,j,\alpha)}\;.
\label{2d-inner-prod-decomp}
\end{eqnarray}
Thus $\delta \Phi^{(y,\beta)}$ produces a variation of the gauge field which is
orthogonal to gauge transformations.

In three space dimensions, we have the decomposition formula
\begin{eqnarray}
\delta A^{(x,j,\alpha)}=
\int L^{(x,j,\alpha)}_{\;\;\;\;\;\;(y,l,\beta)} 
\delta \Phi^{(y,l,\beta)}
+\int{\cal D}^{(x,j,\alpha)}_{\;\;\;\;\;\;(\!(z,\sigma)\!)}
\delta \Omega^{(\!(z,\sigma)\!)} + h^{(x,j,\alpha)}\;,
\label{3d-ym-p-h}
\end{eqnarray}
where
\begin{eqnarray}
{\cal D}^{(x,j,\alpha)}_{\;\;\;\;\;\;(\!(z,\sigma)\!)}
=({\cal D}_{j})^{\alpha}_{\;\;\;\sigma}\delta^{3}(x-z) \;, \nonumber
%\label{3d-ym-normal}
\end{eqnarray}
and
where the vector $h^{(x,j,\alpha)}$
satisfies the harmonicity conditions, which we again call
non-Abelian closed and non-Abelian co-closed, respectively:
\begin{eqnarray}
\int ({\cal D}^{\dagger})^{ (\!(z,\sigma)\!) }_{\;\;\;\;\;\; (x,j,\alpha) }\;
h^{(x,j,\alpha)}=0 \;, \label{3d-ym-closed}
\end{eqnarray}
\begin{eqnarray}
\int 
(L^{\dagger})^{ (y,l,\beta)}_{\;\;\;\;\;\;(x,j, \alpha)} \;
h^{(x,j,\alpha)} =0 \;. 
\label{3d-ym-co-closed}
\end{eqnarray}

If there is a second variation $\delta A^{(x,j,\alpha)}$, of $A^{(x,j,\alpha)}$, we decompose it
as
\begin{eqnarray}
\delta A^{\prime\;(x,j,\alpha)}=
\int L^{(x,j,\alpha)}_{\;\;\;\;\;\;(y,l,\beta)} \delta \Phi^{\prime \;(y,l,\beta)}
+\int{\cal D}^{(x,j,\alpha)}_{\;\;\;\;\;\;(\!(z,\sigma)\!)}
\delta \Omega^{\prime \;(\!(z,\sigma)\!)} + h^{\prime\;(x,j,\alpha)}\;,
\nonumber
\end{eqnarray}

\begin{flushleft}
which yields the decomposition of the inner product:
\end{flushleft}
\begin{eqnarray}
\int \delta A^{\prime\;(x,j,\alpha)}\delta A^{(x,j,\alpha)}=
\int G_{(u,l, \alpha)(v,m,\beta)} \delta \Phi^{\prime \;(u,l,\alpha)}
\delta \Phi^{(v,m,\beta)}
\!&\!+\!&\!  \int (-{\cal D}^{2})_{(\!(z,\gamma )\!)(\!(w,\rho )\!)}
\delta \Omega^{\prime \;(\!(z,\gamma)\!)} 
\delta \Omega^{(\!(w,\rho)\!)} \nonumber \\
\!&\!+\!&\!\int h^{\prime\;(x,j,\alpha)}h^{(x,j,\alpha)}\;.
\label{3d-inner-prod-decomp}
\end{eqnarray}
The quantity $\delta \Phi^{(y,\beta)}$ produces a variation of the gauge field which is
orthogonal to gauge transformations.

We do not prove here that the dimension of the space
of square-integrable
harmonic forms is finite, but this statement seems obvious. For the conditions 
(\ref{2d-ym-closed}) and (\ref{2d-ym-co-closed}) or (\ref{3d-ym-closed}) and (\ref{3d-ym-co-closed}) 
hold if and only
\begin{eqnarray}
\int \diamondsuit^{(x,j,\gamma)}_{\;\;\;\;\;\;(y,k,\omega)}\; h^{(y,k,\omega)}=0\;,
\;\; \diamondsuit = {\cal D} {\cal D}^{\dagger} +L L^{\dagger}  \;. \nonumber
\end{eqnarray}
The operator $\diamondsuit$ is a generalization of the Laplacian on one-forms. This
operator should have a 
pseudo-elliptic self-adjoint extension with a finite number of zero eigenvectors.

We close this section with the following remark. The second
term in the linear
transformation $L$ in two (\ref{2d-ym-exterior-deriv}) or three 
(\ref{3d-ym-exterior-deriv})  space dimensions has the form of a gauge transformation. By
a suitable redefinition 
\begin{eqnarray}
\delta\Omega \longrightarrow \delta\Omega^{\prime}
=\delta \Omega+{\rm i}\frac{1}{{\cal D}^{2}} \;{\cal F}_{12}\; \delta\Phi \;,
\label{2d-ym-redef}
\end{eqnarray}
in two dimensions and
\begin{eqnarray}
\delta\Omega \longrightarrow \delta \Omega^{\prime}=
\delta \Omega+{\rm i}\frac{1}{{\cal D}^{2}} \;\left(\ast{\cal F}\right)\cdot \delta\Phi\;, 
\label{3d-ym-redef}
\end{eqnarray}
in three dimensions, this term can be removed. Then 
\begin{eqnarray}
\delta A^{(x,j,\alpha)}=
\int \epsilon^{jk}{\cal D}^{(x,k,\alpha)}_{\;\;\;\;\;\;(y,\beta)} \delta \Phi^{(y,\beta)}
+\int{\cal D}^{(x,j,\alpha)}_{\;\;\;\;\;\;(\!(z,\sigma)\!)}
\delta \Omega^{\prime\;(\!(z,\sigma)\!)} + h^{(x,j,\alpha)}\;, \label{2d-ym-other-formula}
\end{eqnarray}
\begin{eqnarray}
\delta A^{(x,j,\alpha)}=
\int \epsilon^{jkl}{\cal D}^{(x,k,\alpha)}_{\;\;\;\;\;\;(y,\beta)} 
\delta \Phi^{(y,l,\beta)}
+\int{\cal D}^{(x,j,\alpha)}_{\;\;\;\;\;\;(\!(z,\sigma)\!)}
\delta \Omega^{\prime\;(\!(z,\sigma)\!)} + h^{(x,j,\alpha)}\;,
\label{3d-ym-other-formula}
\end{eqnarray}
in two and three space dimensions, respectively. These relations (\ref{2d-ym-other-formula})
and (\ref{3d-ym-other-formula}), though much simpler, are not as useful from the point of
view of the geometry of orbit space as (\ref{2d-ym-exterior-deriv}) and 
(\ref{3d-ym-exterior-deriv}), since the splitting into gauge transformations
and non-gauge variations is lost. This does not mean
they are not interesting, however, and we shall discuss (\ref{2d-ym-other-formula}) again in Section 8.

\section{Hypersurfaces in connection space}
\setcounter{equation}{0}
\renewcommand{\theequation}{3.\arabic{equation}}

To see the full implication of our decomposition formulas (\ref{2d-ym-p-h}) 
and (\ref{3d-ym-p-h}) for gauge-theory
orbit space, we compare them
to the formula for the coordinates of
a flat space and the coordinates 
of a hypersurface
embedded in this flat space:
\begin{eqnarray}
dy^{a}=e^{a}_{\mu}dx^{\mu}+n^{a}_{H} dw^{H}\;.
\label{embedding} 
\end{eqnarray}
Here the flat space has Cartesian coordinates $y^{a}\;
,\;\;a=1,\dots, n$, the
hypersurface has coordinates $x^{\mu}\;,\;\;\mu=1,\dots, m<n$ 
in a particular
chart and $w^{H}\;,\;\;H=1,\dots, n-m$ are coordinates 
parametrizing
the direction normal to the hypersurface. The analogy we make
is that orbit space $\cal M$ is
a hypersurface
in connection space The hypersurface
coordinates are $\{\Phi\}$ and the normal coordinates are $\{\Omega\}$. The tangent-space basis vectors
$e^{a}_{\mu}$ and the normal vector basis $n^{a}_{H}$ in ${{\rm I}\!{\rm R}}^{n}$ 
correspond in {\it I}$\!\Lambda \!\!\!\!\!-$ to
$L$ and $\cal D$ respectively. Notice that the form of (\ref{2d-ym-p-h}) 
and (\ref{3d-ym-p-h}) implies that the torsion is zero, just as for (\ref{embedding}). Thus, except
at singularities, our coordinates $\{\Phi\}$ are Riemannian.

The coordinates $\{\Phi\}$ we have chosen on gauge orbits are by definition a gauge choice; as is any parametrization
of gauge orbits. They differ from traditional gauge choices, e.g. Landau, Feynman, Coulomb,
Axial, etc., in that variations in the gauge field are covariantly divergenceless. In 
Coulomb gauge $\partial_{j} A^{(x,j,\alpha)}$
in two space dimensions, for example, we may write 
$A^{(x,j,\alpha)}=\epsilon^{jk}
\partial_{k}
\phi_{(x,\alpha)}$. A variation
$\delta \phi_{(x, \alpha)}$ typically produces a change in $A^{(x,j,\alpha)}$ which 
is gauge dependent. This is because
${\cal D}_{j} \epsilon^{jk} \partial_{k}\delta \phi_{(x,\alpha)}\neq 0$, in general. In contrast, 
by virtue of 
(\ref{2d-inner-prod-decomp}) and (\ref{3d-inner-prod-decomp}),
any 
variation in our coordinates
$\{\Phi\}$ is automatically
orthogonal to any gauge transformation in {\it I}$\!\Lambda \!\!\!\!\!-$. 

If boundary conditions can
be chosen with no
harmonic forms, i.e. $h$ satisfying $\diamondsuit h=0$, the analogy between (\ref{2d-ym-p-h}) 
and (\ref{3d-ym-p-h}) and (\ref{embedding}) is complete. This can be done for most gauge
orbits by chosing the boundary conditions appropriately. At 
particular gauge orbits, however, harmonic forms will appear. Harmonic forms in
(\ref{2d-ym-p-h}) 
or (\ref{3d-ym-p-h}) are special gauge transformations. The 
appearance of a harmonic form means that the dimension of the space of
infinitesimal gauge transformations increases by one, and the dimension of infinitesimal
non-gauge variations of the connection decreases by one. Thus harmonic forms appear at values
of $\{\Phi\}$ at the boundary of our coordinate chart. A subset of this boundary
is the Gribov horizon. The 
values of a connected region of $\{\Phi\}$ excluding the boundary are the coordinates of the fundamental
region
of orbit space. 

If the functions describing the embedding of a hypersurface (\ref{embedding}) are sufficiently smooth,
then we may write the differential-geometric formulas
\begin{eqnarray}
\partial_{\mu }e^{a}_{\nu}= \Gamma^{\lambda}_{\mu \nu}\; e^{a}_{\lambda}+{\rm K}^{H}_{\mu \nu}\;n^{a}_{H}\;,
\;\;
\partial_{\mu }n^{a}_{H}= -g^{ \alpha \nu}q_{HJ}{\rm K}^{J}_{\mu \alpha}
\; e^{a}_{\nu}+ \Xi^{I}_{\mu H} n^{a}_{I}
\;, \label{finite-n-hype}
\end{eqnarray}
where $q_{HJ}=\sum_{a}n^{a}_{H}n^{a}_{J}$. We 
recognize $\Gamma^{\lambda}_{\mu \nu} $ as the Riemannian affine connection and
${\rm K}^{H}_{\mu \nu}$ as the second fundamental form or extrinsic curvature. 

If 
we ignore the issue of harmonic forms for the time being, it is sensible to ask what relations 
are analogous to (\ref{finite-n-hype}) for our hypersurface in {\it I}$\!\Lambda \!\!\!\!\!-$. For 
the case of two-dimensional and three-dimensional gauge fields, these are
\begin{eqnarray}
\frac{\delta L^{(u,j,\kappa)}_{\;\;\;\;\;\;(v,\gamma)}}{\delta \Phi^{(x,\alpha)}}
&\!=\!& \int \Gamma^{(z,\rho)}_{(v,\gamma)(x,\alpha)} L^{(u,j,\kappa)}_{\;\;\;\;\;\;(z,\rho)}
+\int {\rm K}^{(\!(t,\sigma)\!)}_{(v,\gamma)(x,\alpha)}  {\cal D}^{(u,j,\kappa)}_{\;\;\;\;\;\;(\!(t,\sigma)\!)}\;,
\nonumber \\
\frac{\delta {\cal D}^{(u,j,\kappa)}_{\;\;\;\;\;\;(\!(s,\tau)\!)}}{\delta \Phi^{(x,\alpha)}}
&\!=\!& -\int G^{ (w,\alpha)(z,\gamma) }
Q_{(\!(s,\tau)\!)(\!(t,\sigma)\!)}{\rm K}^{(\!(t,\sigma)\!)}_{(r,\mu) (w,\alpha)}
\; L^{(u,j,\kappa)}_{\;\;\;\;\;\;(z,\gamma)} \nonumber \\
&\!+\!& \int \Xi^{(\!(t,\sigma)\!)}_{(r,\mu) (\!(s,\tau)\!)} {\cal D}^{(u,j,\kappa)}_{\;\;\;\;\;\;(\!(t,\sigma)\!)}
\;, \label{2d-ym-hype}
\end{eqnarray}
and
\begin{eqnarray}
\frac{\delta L^{(u,j,\kappa)}_{\;\;\;\;\;\;(v,k,\gamma)}}{\delta \Phi^{(x,l,\alpha)}}
&\!=\!& \int \Gamma^{(z,n,\rho)}_{(v,k,\gamma)(x,l,\alpha)} L^{(u,j,\kappa)}_{\;\;\;\;\;\;(z,n,\rho)}
+\int {\rm K}^{(\!(t,\sigma)\!)}_{(v,k,\gamma)(x,l,\alpha)}  {\cal D}^{(u,j,\kappa)}_{\;\;\;\;\;\;(\!(t,\sigma)\!)}\;,
\nonumber \\
\frac{\delta {\cal D}^{(u,j,\kappa)}_{\;\;\;\;\;\;(\!(s,\tau)\!)}}{\delta \Phi^{(x,l,\alpha)}}
&\!=\!& -\int G^{(w,m,\beta) (z,k,\gamma) }Q_{(\!(s,\tau)\!)(\!(t,\sigma)\!)}
{\rm K}^{(\!(t,\sigma)\!)}_{(x,l,\alpha) (w,m,\beta)}
\; L^{(u,j,\kappa)}_{\;\;\;\;\;\;(z,k,\gamma)} \nonumber \\
&\!+\!& \int \Xi^{(\!(t,\sigma)\!)}_{(x,l,\alpha) (\!(s,\tau)\!)} {\cal D}^{(u,j,\kappa)}_{\;\;\;\;\;\;(\!(t,\sigma)\!)}
\;, \label{3d-ym-hype}
\end{eqnarray}
respectively, where 
\begin{eqnarray}
Q _{(\!(s,\tau)\!) (\!(t,\sigma  )\!) }   =\int {\cal D}^{(u,j,{\kappa})}_{\;\;\;\;\;\;(\!(s,\tau)\!)} 
{\cal D}^{(u,j,{\kappa})}_{\;\;\;\;\;\;(\!(t,\sigma)\!)}=(-{\cal D}^{2})_{(\!(s,\tau)\!) (\!(t,\sigma  )\!) }   \;.
\label{normal-metric}
\end{eqnarray}

\section{Gauge-invariant or gauge-covariant?}
\setcounter{equation}{0}
\renewcommand{\theequation}{4.\arabic{equation}}

We have not yet worked explicitly with the components of $\{\Phi\}$, but only its variation. An explicit 
expression for these components is given in Section 8, though only for the (2+1)-dimensional case. We have
claimed that $\{\Phi\}$ constitutes a set of gauge-invariant coordinates (if
we chose the domain of $\{\Phi\}$ to be the fundamental region) on orbit space. Though this claim is
very easy to justify, it is so crucial to the interpretation of this paper that we felt it necessary
to provide the justification.

Note that the metric tensors (\ref{2d-metric-tensor}) and (\ref{3d-metric-tensor}) are
formally gauge-covariant expressions, and not gauge-invariant. Thus $\delta \Phi$ is gauge-covariant
as well. Under a gauge transformation $g(x)\in$ SU($N$) with adjoint representation $D_{\rm ADJ}(g(x))$,
\begin{eqnarray}
\delta\Phi^{(x,\alpha)} \rightarrow \delta\Phi^{\prime\;(x,\alpha)}=
D_{\rm ADJ}(g(x))^{\alpha}_{\;\;\;\beta}\;
\delta\Phi^{(x,\beta)} \;,\label{2d-gauge-variation}
\end{eqnarray}
in 2+1 dimensions and
\begin{eqnarray}
\delta\Phi^{(x,k,\alpha)} \rightarrow \delta\Phi^{\prime\;(x,k,\alpha)}=
D_{\rm ADJ}(g(x))^{\alpha}_{\;\;\;\beta}\;
\delta\Phi^{(x,k,\beta)} \;,\label{3d-gauge-variation}
\end{eqnarray}
in 3+1 dimensions. To properly define $\{\Phi\}$, however, it is necessary to integrate these 
expressions. To do this, a particular orbit, chosen by taking some particular connection $A^{(x,j,\alpha)}$
is identified with a particular
choice of $\{\Phi\}$, say $\{\phi\}$. This choice is a boundary condition on the first-order
functional differential equation
\begin{eqnarray}
\delta A^{(x,j,\alpha)} =\int L^{(x,j,\alpha)}_{\;\;\;\;\;\;(y,\beta)}\;\delta \Phi^{(y,\beta)}\;, 
\label{2d-funct-diff-eq}
\end{eqnarray}
in two space dimensions or
\begin{eqnarray}
\delta A^{(x,j,\alpha)} =\int L^{(x,j,\alpha)}_{\;\;\;\;\;\;(y,l,\beta)}\;\delta \Phi^{(y,l,\beta)}\;, 
\label{3d-funct-diff-eq}
\end{eqnarray}
in three space dimensions. These variations remain
on the hypersurface. A gauge transformation changes the 
boundary condition, and thereby changes the hypersurface. Therefore, we 
can assign a choice of $\{\Phi\}$ and the gauge field (in the fundamental region) to each orbit
(except on a set of measure zero). This is why our
coordinates are gauge-invariant, not gauge-covariant. 

It is important that ${\cal D}^{2}$ not have zero modes. We can achieve this by putting the system
in a finite volume, say a two-dimensional disk or three-dimensional ball. At the boundary, we set
all components of $\{\phi\}$ and $\{\delta \Phi \}$ equal to zero. The gauge connection is defined by
integrating
(\ref{2d-funct-diff-eq}) or (\ref{3d-funct-diff-eq}). Then the operator manipulations we have
performed remain legitimate.

\section{The extrinsic and intrinsic curvatures of the hypersurface}
\setcounter{equation}{0}
\renewcommand{\theequation}{5.\arabic{equation}}

We shall next use equations (\ref{2d-ym-hype}) and (\ref{3d-ym-hype}) 
to obtain the second fundamental form and the intrinsic curvature of orbit space. The curvature tensor
was first discussed many years ago \cite{singer, bab-via2}. Knowledge of the intrinsic curvature
can give important global information relevant to the spectrum of the Laplacian \cite{dav-chav}.

For the case of a hypersurface in finite-dimensional space, the equations (\ref{finite-n-hype}) imply that
\begin{eqnarray}
{\rm K}_{H,\mu \nu}=q_{HJ} {\rm K}^{J}_{\mu \nu}=-\sum_{a} e^{a}_{\mu} \partial_{\nu} n^{a}_{H}\;.  \label{finite-n-sec-fund-form}
\end{eqnarray}
The curvature tensor is given in terms of the second fundamental form by Gauss' formula:
\begin{eqnarray}
R_{\alpha \beta \mu \nu}= q^{HJ}
[ {\rm K}_{H,\alpha \mu} {\rm K}_{J,\beta \nu}- {\rm K}_{H,\alpha \beta} {\rm K}_{J,\mu \nu}] \;,
\label{finite-n-riemann}
\end{eqnarray}
where $q^{HJ}$ is the inverse of $q_{HJ}$.

We first consider the second fundamental form for (2+1)-dimensional gauge theories. The 
generalization of (\ref{finite-n-sec-fund-form}) to the hypersurface in connection space is
\begin{eqnarray}
{\rm K}_{(\!(z,\lambda)\!), (x,\alpha) (y,\beta)}
\!&\!=\!&\! -\int \; L^{(u,j,\gamma)}_{\;\;\;\;\;\;(x,\alpha)}
\frac{\delta {\cal D}^{(u,j,\gamma)}_{\;\;\;\;\;\;(\!(z,\lambda)\!)}}{\delta \Phi^{(y,\beta)} } 
=-\int \; L^{(u,j,\gamma)}_{\;\;\;\;\;\;(x,\alpha)} L^{(v,k,\sigma)}_{\;\;\;\;\;\;(y,\beta)}
\frac{\delta {\cal D}^{(u,j,\gamma)}_{\;\;\;\;\;\;(\!(z,\lambda)\!)}}{\delta A^{(v,k,\sigma)} } 
\nonumber \\
\!&\!=\!&\!-f_{\lambda \gamma \sigma}
L^{(z,j,\gamma)}_{\;\;\;\;\;\;(x,\alpha)}
L^{(z,j,\sigma)}_{\;\;\;\;\;\;(y,\beta)} \;,\label{2d-ym-sec-fund-form}
\end{eqnarray}
where in the last expression the sum is implicit on $j$, $\gamma$ and $\sigma$. For 3+1 dimensions
\begin{eqnarray}
{\rm K}_{(\!(z,\lambda)\!), (x,l,\alpha) (y,m,\beta)}=
-f_{\lambda \gamma \sigma}
L^{(z,j,\gamma)}_{\;\;\;\;\;\;(x,l,\alpha)}
L^{(z,j,\sigma)}_{\;\;\;\;\;\;(y,m,\beta)} \;.\label{3d-ym-sec-fund-form}
\end{eqnarray}
Notice that there is no integration in the final form of either of these expressions.

The curvature
(\ref{finite-n-riemann}) generalized to (2+1)-dimensional gauge theories is 
\begin{eqnarray}
R_{(x,\alpha)(y,\beta)(u,\mu)(v,\nu)}
\!&\!=\!&\!  \int \; \left[ \left( \frac{1}{-{\cal D}^{2}} \right)^{\lambda \rho}\delta^{2}(z-w)\right]\;
\left[ {\rm K}_{(\!(z,\lambda )\!),(x,\alpha)(u,\mu)}  {\rm K}_{(\!(w,\rho)\!),(y,\beta)(v,\nu)} 
\right. \nonumber \\
\!&\!-\!&\!
\left.
{\rm K}_{(\!(z,\lambda)\!),(x,\alpha)(y,\beta)}  {\rm K}_{(\!(w,\rho)\!),(u,\mu)(v,\nu)} 
\right] \;. \label{babelon-viallet} 
\end{eqnarray}
An expression similar to (\ref{babelon-viallet}) can 
be found in reference \cite{bab-via2}, but the second fundamental
form was 
not explicitly 
determined there. Inserting (\ref{2d-ym-sec-fund-form}) into (\ref{babelon-viallet}) yields
\begin{eqnarray}
R_{(x,\alpha)(y,\beta)(u,\mu)(v,\nu)}
\!&\!=\!&\!\int \; \left[\left( \frac{1}{-{\cal D}^{2}} \right)^{\lambda \rho}\delta^{2}(z-w) \right]\;
f_{\lambda \gamma \sigma}f_{\rho \kappa \tau} \nonumber\\
\!&\!\times\!&\!
\left[ 
L^{(z,j,\gamma)}_{\;\;\;\;\;\;(x,\alpha)}
L^{(z,j,\sigma)}_{\;\;\;\;\;\;(u,\mu)}
L^{(w,k,\kappa)}_{\;\;\;\;\;\;(y,\beta)}
L^{(w,k,\tau)}_{\;\;\;\;\;\;(v,\nu)} \right. 
\;\;\;\;\;\;\;\;\;\;\;\;\;\;\;\;\;\;\;\;\;\;\;\;\;\;\;\;\;\;\;\;\;\; 
\nonumber
\end{eqnarray}
\begin{eqnarray}
\;\;\;\;\;\;\;\;\;\;\;\;\;\;\;\;\;\;\;\;\;\;\;\;\;\;\;\;\;\;\;\;\;\; 
\left.
-L^{(z,j,\gamma)}_{\;\;\;\;\;\;(x,\alpha)}
L^{(z,j,\sigma)}_{\;\;\;\;\;\;(y,\beta)}
L^{(w,k,\kappa)}_{\;\;\;\;\;\;(u,\mu)}
L^{(w,k,\tau)}_{\;\;\;\;\;\;(v,\nu)}
\right]\;.
\label{2d-ym-riemann}
\end{eqnarray}
The 
corresponding expression for (3+1)-dimensional gauge theories is
\begin{eqnarray}
R_{(x,l,\alpha)(y,m,\beta)(u,p,\mu)(v,r,\nu)}
=\int \; \left[ \left( \frac{1}{-{\cal D}^{2}} \right)^{\lambda \rho}\delta^{3}(z-w) \right]\;
f_{\lambda \gamma \sigma}f_{\rho \kappa \tau} \nonumber
\end{eqnarray}
\begin{eqnarray}
\times
\left[ 
L^{(z,j,\gamma)}_{\;\;\;\;\;\;(x,l,\alpha)}
L^{(z,j,\sigma)}_{\;\;\;\;\;\;(u,p,\mu)}
L^{(w,k,\kappa)}_{\;\;\;\;\;\;(y,m,\beta)}
L^{(w,k,\tau)}_{\;\;\;\;\;\;(v,r,\nu)} \right.  
\;\;\;\;\;\;\;\;\;\;\;\;\;\;\;\;\;\;\;\;\;\;\;\;\;\;\;\;\;\;\;\;\;\; 
\nonumber
\end{eqnarray}
\begin{eqnarray}
\;\;\;\;\;\;\;\;\;\;\;\;\;\;\;\;\;\;\;\;\;\;\;\;\;\;\;\;\;\;\;\;\;\;
\left.-
L^{(z,j,\gamma)}_{\;\;\;\;\;\;(x,l,\alpha)}
L^{(z,j,\sigma)}_{\;\;\;\;\;\;(y,m,\beta)}
L^{(w,k,\kappa)}_{\;\;\;\;\;\;(u,p,\mu)}
L^{(w,k,\tau)}_{\;\;\;\;\;\;(v,r,\nu)}
\right]\;,
\label{3d-ym-riemann}
\end{eqnarray}
where $l,m,p,r=1,2$. These expressions for the curvature
are evidently well-defined with no regularization.  This is no longer the case
for the contractions of the curvature tensor, namely the Ricci and scalar curvatures.

\section{Sectional, Ricci and scalar curvatures for Yang-Mills theory in arbitrary dimensions}
\setcounter{equation}{0}
\renewcommand{\theequation}{6.\arabic{equation}}

We will next discuss the curvature tensor of orbit space 
and its contractions for an arbitrary number of space
dimensions D (so space-time has D+1 dimensions). We anticipate that our expressions for
the Ricci and scalar curvatures may be used
to calculate these quantities in dimensional regularization. We hope to perform
these calculations near particular gauge-field backgrounds in a future publication. We should
mention that Singer discussed a
calculation of the Ricci curvature using zeta-function regularization \cite{singer}, though some
details were evidently never published.

The non-Abelian exactness condition in D space dimensions, generalizing (\ref{2d-ym-exact}) and
(\ref{3d-ym-exact}), is
\begin{eqnarray}
\delta A^{(x,j,\alpha)}= L^{(x,j,\alpha)}_{\;\;\;\;\;\;(y,l_{1} l_{2} \cdots l_{{\rm D}-2}, \beta)}
\;  \delta \Phi^{(y,l_{1}l_{2} \cdots l_{{\rm D}-2} ,\beta)}   \;, \label{Dd-ym-exact}
\end{eqnarray}
where the mapping from the hypersurface to connection space in any space dimension D, generalizing
(\ref{2d-ym-exterior-deriv}) and
(\ref{3d-ym-exterior-deriv}) is
\begin{eqnarray}
L^{(x,j,\alpha)}_{\;\;\;\;\;\;(y,l_{1}l_{2}\cdots l_{{\rm D}-2},\beta)}
=\frac{\delta A^{(x,j,\alpha)}}{\delta \Phi^{(y,l_{1} l_{2} \cdots l_{{\rm D}-2} ,\beta)}}
\;\;\;\;\;\;\;\;\;\;\;\;\;\;\;\;\;\;\;\;\;\;\;\;\;\;\;\;\;\;\;\;\;\;\;\;\;\;\;\;\;
\nonumber 
\end{eqnarray}
\begin{eqnarray}
\;\;\;\;\;\;\;\;\;\;\;\;\;\;=\left[ \epsilon^{jkl_{1}l_{2}\cdots l_{{\rm D}-2}}{\cal D}_{k}+{\rm i}
{\cal D}_{j} \frac{1}{{\cal D}^{2}} (\ast {\cal F})^{(x,l_{1}l_{2}\cdots l_{{\rm D}-2})}
\right]^{\alpha}_{\;\;\beta} \delta^{\rm D}(x-y)\;,
\label{Dd-exterior-deriv}
\end{eqnarray}
The variation $\delta \Phi^{(y,l_{1}l_{2} \cdots l_{{\rm D}-2} ,\beta)}$ is fully antisymmetric
in the space indices $l_{1},l_{2}, \dots, l_{{\rm D}-2}$. As we would expect, the field dual to
a vector-gauge field is an antisymmetric-tensor gauge field of rank D$-2$ (in d space-time dimensions,
this is d-3). The infinitesimal
dual gauge transformations
are the zero modes of the mapping $L$.

The metric tensor for D$>2$ dimensions is not
\begin{eqnarray}
G_{(u,\{p\},\mu) (v,\{r\},\nu) }= \int 
L^{(x, j,\alpha) }_{\;\;\;\;\;\;\; (u,\{p\},\mu)  }
L^{ (x, j,\alpha)  }_{\;\;\;\;\;\;\; (v,\{r\},\nu)  } \;,
\label{unprojected-metric}
\end{eqnarray}
but rather
\begin{eqnarray}
G_{XY}= \int (P_{\rm dual})_{X}^{\;\;\; (u,\{p\},\mu) }(P_{\rm dual})_{X}^{\;\;\;(v,\{r\},\nu)} 
L^{(x, j,\alpha) }_{\;\;\;\;\;\;\;(u,\{p\},\mu)  }
L^{ (x, j,\alpha)  }_{\;\;\;\;\;\;\;(v,\{r\},\nu)  } \;,
\label{projected-metric}
\end{eqnarray}
where
$(P_{\rm dual})_{X}^{\;\;\;(x,\{l\},\alpha)}$ is the projection operator
which projects out the zero modes of $L$. Here the indices
$X,Y$ are those of the subspace with these zero modes removed. This solution of the dual-gauge-invariance
problem is more
abstract than used earlier in this paper, but is also more convenient for the goal of determining the 
Ricci and scalar curvatures for arbitrary space dimensions.

Before projecting out the zero modes of $L$, the 
curvature tensor, determined by the methods of the last section is
\begin{eqnarray}
R_{(x,\{l\},\alpha)(y,\{m\},\beta)(u,\{p\},\mu)(v,\{r\},\nu)}
=\int \; \left[ \left( \frac{1}{-{\cal D}^{2}} \right)^{\lambda \rho}\delta^{\rm D}(z-w) \right]\;
f_{\lambda \gamma \sigma}f_{\rho \kappa \tau} \nonumber
\end{eqnarray}
\begin{eqnarray}
\times
\left[ 
L^{(z,j,\gamma)}_{\;\;\;\;\;\;(x,\{l\},\alpha)}
L^{(z,j,\sigma)}_{\;\;\;\;\;\;(u,\{p\},\mu)}
L^{(w,k,\kappa)}_{\;\;\;\;\;\;(y,\{m\},\beta)}
L^{(w,k,\tau)}_{\;\;\;\;\;\;(v,\{r\},\nu)} \right.  
\;\;\;\;\;\;\;\;\;\;\;\;\;\;\;\;\;\;\;\;\;\;\;\;\;\;\;\;\;\;\;\;\;\; 
\nonumber
\end{eqnarray}
\begin{eqnarray}
\;\;\;\;\;\;\;\;\;\;\;\;\;\;\;\;\;\;\;\;\;\;\;\;\;\;\;\;\;\;\;\;\;\;
\left.-
L^{(z,j,\gamma)}_{\;\;\;\;\;\;(x,\{l\},\alpha)}
L^{(z,j,\sigma)}_{\;\;\;\;\;\;(y,\{m\},\beta)}
L^{(w,k,\kappa)}_{\;\;\;\;\;\;(u,\{p\},\mu)}
L^{(w,k,\tau)}_{\;\;\;\;\;\;(v,\{r\},\nu)}
\right]\;,
\label{Dd-ym-riemann}
\end{eqnarray}
where $\{l\}$ is an abbreviation for $l_{1}, l_{2},\dots,l_{{\rm D}-2}$, etc. Again, some of the components
of (\ref{Dd-ym-riemann}) should be removed; this is because of the dual gauge invariance. We do
this using projection operators, instead of restricting indices. The actual 
curvature ${\cal R}_{XYUV}$ for D$>2$ is given by 
\begin{eqnarray}
{\cal R}_{XYUV}\!&\!=\!&\!\int (P_{\rm dual})_{X}^{\;\;\;(x,\{l\},\alpha)}
(P_{\rm dual})_{Y}^{\;\;\;(y,\{m\},\beta)} (P_{\rm dual})_{U}^{\;\;\;(u,\{p\},\mu)}
(P_{\rm dual})_{V}^{\;\;\;(v,\{r\},\nu)} 
\nonumber \\
\!&\!\times\!&\!
R_{(x,\{l\},\alpha)(y,\{m\},\beta)(u,\{p\},\mu)(v,\{r\},\nu)}\;. \label{corrected-riemann}
\end{eqnarray}

The sectional curvature between (real) vectors $\xi^{X}$, $\zeta^{X}$ is
\begin{eqnarray}
\frac{\int {\cal R}_{XYUV}\xi^{X}\xi^{Y}\zeta^{U}\zeta^{V}}{ \left< \xi \vert \zeta\right>^{2} }
\;, \label{sectional}
\end{eqnarray}
where
\begin{eqnarray}
\left< \xi \vert \zeta\right>= \int G_{(x,\{l\}, \alpha)(y,\{m\}, \beta)}
(P_{\rm dual})_{X}^{\;\;\;(x,\{l\},\alpha)} 
(P_{\rm dual})_{Y}^{\;\;\;(x,\{m\},\beta)} \xi^{X}\zeta^{Y}\;. \label{proj-in-prod} 
\end{eqnarray}
Due to the antisymmetry of the group structure coefficients $\{f\}$, the second term of the curvature 
tensor does
not contribute to the numerator of (\ref{sectional}); hence the sectional curvature is non-negative.

We introduce a second projection operator
\begin{eqnarray}
P^{(x,j,\alpha)(y,k,\beta)}=\left[\delta_{jk}\delta^{\alpha \beta}
-{\cal D}_{j} \frac{1}{{\cal D}^{2}} {\cal D}_{k}
\right]\delta^{\rm D}(x-y)\;, \label{project}
\end{eqnarray}
which has the same form as the metric discussed in the introduction (\ref{first-metric-tensor}). This projects out
ordinary Yang-Mills gauge transformations.

To obtain the Ricci curvature, we must contract (i.e. multiply and carry out the sum and integration
over repeated indices) the curvature tensor (\ref{corrected-riemann}) with the inverse metric. This is
\begin{eqnarray}
G^{UV}= \int (P_{\rm dual})_{U}^{\;\;\; (u,\{p\},\mu) }(P_{\rm dual})_{V}^{\;\;\;(v,\{r\},\nu)} 
(L^{-1})^{(u,\{p\},\mu)}_{\;\;\;\;\;\;\;(x, j,\alpha)}
(L^{-1})^{(v,\{r\},\nu)}_{\;\;\;\;\;\;\;(y, k,\beta)} P^{(x,j,\alpha)(y,k,\beta)}\;.
\label{projected-inverse-metric}
\end{eqnarray}
We have inserted $P$ and $P_{\rm dual}$ into (\ref{projected-inverse-metric}) to
keep this expression well-defined. 

We are finally ready to write down the Ricci tensor. We contract the curvature tensor (\ref{corrected-riemann})
with (\ref{projected-inverse-metric}) to obtain
\begin{eqnarray}
{\rm Ric}_{XY}=\int (P_{\rm dual})_{X}^{\;\;\;(x,\{l\},\alpha)}
(P_{\rm dual})_{Y}^{\;\;\;(y,\{m\},\beta)} {\rm Ric}_{(x,\{l\},\alpha)(y,\{m\},\beta)}
\;, \label{ricci-1}
\end{eqnarray}
where
\begin{eqnarray}
{\rm Ric}_{(x,\{l\},\alpha)(y,\{m\},\beta)} \!&=\!&\! \int \; 
\left[ \left( \frac{1}{-{\cal D}^{2}} \right)^{\lambda \rho}\delta^{\rm D}(z-w) \right]\;
f_{\lambda \gamma \sigma}f_{\rho \kappa \tau} \nonumber \\
\!&\!\times\!&\! L^{(z,j,\gamma)}_{\;\;\;\;\;\;(x,\{l\},\alpha)}
L^{(w,k,\kappa)}_{\;\;\;\;\;\;(y,\{m\},\beta)}
P_{(z,j,\sigma)(w,k,\tau)} \;. \label{ricci-2}
\end{eqnarray}
Notice that the Ricci curvature is formally non-negative. Once this expression is regularized, it will
be important to see if a positive lower bound exists on its spectrum.

The scalar curvature is obtained by a further contraction of the Ricci curvature with the inverse metric. It is
comparatively simple, containing no factors of $P_{\rm dual}$:
\begin{eqnarray}
{\rm R}=\int \; \left[ \left( \frac{1}{-{\cal D}^{2}} \right)^{\lambda \rho}\delta^{\rm D}(z-w) \right]\;
f_{\lambda \gamma \sigma}f_{\rho \kappa \tau} 
P_{(z,j,\gamma)(w,k,\kappa)} P_{(z,j,\sigma)(w,k,\tau)} \;. \label{scalar-curvature}
\end{eqnarray}

As mentioned at the beginning at the section, we hope to study these quantities further in a later
publication. Our intention is to expand the Green's functions in (\ref{ricci-2})
and (\ref{scalar-curvature}) around particular backgrounds with space dimension $\rm D$ close to
an integer value.

\section{The Yang-Mills Hamiltonian}
\setcounter{equation}{0}
\renewcommand{\theequation}{7.\arabic{equation}}

In this section, we will examine the kinetic term of the Hamiltonian of 
non-Abelian gauge theories
in terms of the gauge-invariant coordinates $\{\Phi\}$ and consider the functional form of the measure. 

The Yang-Mills Hamiltonian operator is 
\begin{eqnarray}
H=T+U\;, \;\;\ T=-\frac{e_{0}^{2}}{2}\int \frac{\delta^{2}}{\delta A^{(x,j,\alpha)}\delta A^{(x,j,\alpha)}}
\;,\;\; U=\frac{1}{2e_{0}^{2}}\int (F_{jk})^{2} \;, \label{ym-hamiltonian}
\end{eqnarray}
and wave functions depend only on $\Phi$ by Gauss' law. The coupling constant $e_{0}$ has
engineering dimension $(4-d)/2$, where $d$ is the dimension of space-time. 

The kinetic term in 2+1 dimensions may be written as
the Laplacian (we use $e$ rather than $e_{0}$, since the theory is finite) 
\begin{eqnarray}
T
=-\frac{e^{2}}{2}\int \;
\frac{1}{\sqrt{{\rm det}G}}\;
\frac{\delta}{\delta\Phi^{(y,\beta)}}\; {\sqrt{{\rm det}G}}\;G^{(y,\beta)(z,\gamma)}\;
\frac{\delta}{\delta\Phi^{(z,\gamma)}}\;,
\label{2d-phi-ym-kinetic}
\end{eqnarray}
where the $G_{(x,\alpha)(y,\beta)}$ and $G^{(x,\alpha)(y,\beta)}$ 
are the component of the metric tensor (\ref{2d-metric-tensor}) and its inverse, respectively, on 
the hypersurface. In 3+1 dimensions
\begin{eqnarray}
T
=-\frac{e_{0}^{2}}{2}\int \;
\frac{1}{\sqrt{{\rm det}G}}\;
\frac{\delta}{\delta\Phi^{(y,l,\beta)}}\; {\sqrt{{\rm det}G}}\;G^{(y,l,\beta)(z,m,\gamma)}\;
\frac{\delta}{\delta\Phi^{(z,m,\gamma)}}\;,
\label{3d-phi-ym-kinetic}
\end{eqnarray}
with $l,m=1,2$. Alternative expressions can be given in by chosing different dual gauge fixings.

The 
inner product on Schr\"{o}dinger wave functionals is the functional
integral
\begin{eqnarray}
\left< \Psi_{1} \vert \Psi_{2} \right>
=\int [d\Phi]  \;{\sqrt {{\rm det}G}}\; {\overline {\Psi}}_{1}[\Phi] \Psi_{2}[\Phi]\;.
\label{inner-product}
\end{eqnarray}
It is the measure factor in this inner product which is responsible for the appearance of the mass
gap \cite{kar-nair,kar-kim-nair}, as we discuss in Section 10.

Strictly speaking the Hamiltonian may have another term equal to $\frac{{\rm R}}{6}$ where R is the
scalar curvature (\ref{scalar-curvature}). It was pointed out by Gawedzki \cite{gawed}
that such a term can be present. If one begins with (\ref{ym-hamiltonian}) and (\ref{2d-phi-ym-kinetic}) or
(\ref{2d-phi-ym-kinetic}) and uses the Trotter product 
formula \cite{simon} to determine the action principle one finds a Euclidean action
\begin{eqnarray}
S=\int \frac{1}{2e_{0}^{2}} {\dot \Phi}^{T} G {\dot \Phi}+ \int d\tau \frac{{\rm R}}{6} 
+\int \frac{1}{4e_{0}^{2}} {\rm Tr} F_{ij}F_{ij}
\;, \nonumber
\end{eqnarray}
where $T$ denotes the transpose, $\tau$ denotes Euclidean time and the dot denotes time-differentiation. Removal
of the scalar-curvature term from the action will force this term into the Hamiltonian. It is not yet clear
how important this term is; it may be irrelevant (in the renormalization group sense) in which case
it may be dropped. If the term is not irrelevant, however, it may be essential for 
a Lorentz-invariant spectrum. We leave this question open in the remaining of this paper, though we
believe that it can be answered by the calculation we have proposed at the end of the last section.

\section{Solving for $\{\Phi\}$ in $2+1$ dimensions}
\setcounter{equation}{0}
\renewcommand{\theequation}{8.\arabic{equation}}

A drawback of our formalism is that we can only write the variation of the gauge field in terms of the
variation of $\{\Phi\}$. We would like an expression for $A^{(x,j,\alpha)}$ in terms of $\Phi^{(y,\lambda)}$. Though
we have not yet solved this problem, we have done the reverse by finding an expression for $\Phi^{(y,\lambda)}$
in terms of $A^{(x,j,\alpha)}$ in two space dimensions. 

We define holomorphic coordinates $z=x^{1}+{\rm i}x^{2}$,
${\bar z}=x^{1}-{\rm i}x^{2}$ and their derivatives $\partial=\frac{1}{2}( \partial_{1}-{\rm i}\partial_{2})$,
${\bar{\partial}}=\frac{1}{2}(\partial_{1}+{\rm i}\partial_{2})$ (we are no longer using
the letter $z$ to represent a Cartesian coordinate of space, as we did in earlier sections). The gauge field in
the fundamental representation has the components 
$A^{(z,{\bar z},\alpha)}
=\frac{1}{2}(A^{(x,1,\alpha)}-{\rm i}A^{(x,2,\alpha)})$, ${\bar A}^{(z,{\bar z},\alpha)}
=\frac{1}{2}(A^{(x,1,\alpha)}+{\rm i}A^{(x,2,\alpha)})$, in these coordinates. We also define 
${\cal D}=\frac{1}{2}( {\cal D}_{1} - {\rm i} {\cal D}_{2} )$,
${\overline {\cal D}}=\frac{1}{2}( {\cal D}_{1}+{\rm i}{\cal D}_{2} )$. 

In holomorphic coordinates, equation (\ref{2d-ym-p-h}) becomes
\begin{eqnarray}
\delta A^{( z,{\bar z},\alpha )}=
{\rm i}\left[ {\cal D}\left(1+\frac{1}{ {\cal D}{\overline {\cal D}} +{\overline {\cal D}}{\cal D}  }
[{\cal D},{\overline {\cal D}}]\right)\right]^{\alpha}_{\;\;\;\beta} \delta \Phi^{(z,{\bar z}, \beta)} 
\!&\!+\!&\!{\cal D}^{\alpha}_{\;\;\;\beta} \delta \Omega^{(z,{\bar z}, \beta)} 
+h^{(z,{\bar z},\alpha )} \;, \nonumber\\
\delta {\bar A}^{(z,{\bar z},\alpha)}=
-{\rm i} \left[{\overline {\cal D}}\left(1-\frac{1}{ {\cal D}{\overline {\cal D}} +{\overline {\cal D}}{\cal D}  }
[{\cal D},{\overline {\cal D}}]\right)\right]^{\alpha}_{\;\;\;\beta} \delta \Phi^{(z,{\bar z}, \beta)}  
\!&\!+\!&\!{\overline {\cal D}}^{\alpha}_{\;\;\;\beta} \delta \Omega^{(z,{\bar z}, \beta)}   \nonumber \\
\!&\!+\!&\!{\bar h}^{(z,{\bar z},\alpha )}\;.
\label{holomorphic}
\end{eqnarray}
If there are no harmonic forms, the equations (\ref{holomorphic}) imply the remarkable relation
\begin{eqnarray}
\delta \Phi = -\frac{\rm i}{2} \;{\cal D}^{-1}\delta A\;+\;\frac{\rm i}{2}\;{\overline {\cal D}}^{-1}\delta {\bar A}\;.
\label{remarkable}
\end{eqnarray}

We would like to integrate (\ref{remarkable}). Notice that $\delta \Phi$ can be regarded as a one-form
on infinite-dimensional connection space {\it I}$\!\Lambda \!\!\!\!\!-$. Furthermore, this one-form is closed:
\begin{eqnarray}
\frac{\delta}{\delta A^{(z,{\bar z},\alpha)}} \frac{\rm i}{2}\;{\overline {\cal D}}^{-1}-
\frac{\delta}{\delta {\bar A}^{(z,{\bar z},\beta)} } (-)\frac{\rm i}{2}\;{\cal D}^{-1}=0+0=0 \;,\label{functional-closed}
\end{eqnarray}
except at values of $A,{\bar A}$ where ${\cal D}$ or ${\overline {\cal D}}$ has a zero 
mode. Therefore, if some particular choice of $\{\Phi\}$, namely $\{\phi\}$ is identified with the 
gauge field $\{B\}, \{ {\bar B}\}$, we have
\begin{eqnarray}
\Phi=\phi+\int_{\phi}^{\Phi} \;d \Phi^{\prime}= 
\phi+\int_{B, {\bar B}}^{A, {\bar A}}\; 
\left[-\frac{\rm i}{2} \;\frac{1}{{\cal D}^{\prime}}\; dA^{\prime}\;+
\;\frac{\rm i}{2}\; \frac{1}{{\overline {\cal D}}^{\prime}}  \;d{\bar A}^{\prime} \right]\;.
\label{integral-formula}
\end{eqnarray}
By virtue of (\ref{functional-closed}), this integral is independent of the path of
integration chosen in {\it I}$\!\Lambda \!\!\!\!\!-$, provided the path does not encounter
any values of $A^{\prime}, {\bar A}^{\prime}$ where either ${\cal D}^{\prime}$ or
${\overline {\cal D}}^{\prime}$ has a zero eigenvalue. Furthermore, if the path does not encounter such 
singularities, $\diamondsuit$ will
not have a zero eigenvalue and there are no harmonic forms to worry about.  Therefore, the question
of whether ${\cal D}b=0$ or ${\overline {\cal D}}{\bar b}=0$ has a solution is important. In a finite volume, these
solutions will not exist for most gauge fields \cite{foot2}.

An explicit
realization of (\ref{integral-formula}) is obtained by taking
\begin{eqnarray}
A^{\prime}\!&\!=\!&\!A(x,T)=T[A(x)-B(x)]+B(x)\;, \nonumber \\
{\bar A}^{\prime}\!&\!=\!&\!{\bar A}(x,T)=T[{\bar A}(x)-{\bar B}(x)]+{\bar B}(x)\;, \nonumber
%\label{path-choice}
\end{eqnarray}
for $0\le T\le 1$. Thus (\ref{integral-formula}) becomes, at least formally 
\begin{eqnarray}
\Phi=\phi&\!-\!&\frac{\rm i}{2}\int_{0}^{1} \;d T
\left[ \; \frac{1}{\partial-{\rm i}T( {\cal A}-{{\cal B}})-{\rm i}{{\cal B}} } \;(A-B) \right. \nonumber \\
\!&\!-\!&\!
\left. \frac{1}{{\bar \partial}-{\rm i}T({\bar {\cal A}}-{\bar {\cal B}})-{\rm i}{\bar{\cal B}} }
\; ({\bar A}-{\bar B}) \right]\;,
\label{explicit-integral-formula}
\end{eqnarray}
which should be well-defined, provided no zero modes exist for 
${\partial-{\rm i}T( {\cal A}-{{\cal B}})-{\rm i}{{\cal B}} }$ or 
${{\bar \partial}-{\rm i}T({\bar {\cal A}}-{\bar {\cal B}})-{\rm i}{\bar{\cal B}} }$ for any
$T$ in the closed interval $[0,1]$.

We can make the conclusions above more transparent by making the following redefinition of 
$\delta \Omega^{(z,{\bar z}, \beta)} $
in the non-Abelian decomposition formula (\ref{holomorphic})
\begin{eqnarray}
\delta \Omega^{( z,{\bar z},\alpha )} \longrightarrow \delta \Omega^{\prime\;(z,{\bar z}, \beta)} 
=\delta \Omega^{( z,{\bar z},\alpha )} +
{\rm i} \left(
\frac{1}{ {\cal D}{\overline {\cal D}} +{\overline {\cal D}}{\cal D}  }
[{\cal D},{\overline {\cal D}}]\right)^{\alpha}_{\;\;\;\beta} \delta \Phi^{(z,{\bar z}, \beta)} \;.
\label{absorb}
\end{eqnarray}
With this redefinition (\ref{holomorphic}) becomes
\begin{eqnarray}
\delta A^{( z,{\bar z},\alpha )}=
{\rm i}{{\cal D}}^{\alpha}_{\;\;\;\beta} \delta \Phi^{(z,{\bar z}, \beta)} 
+{\cal D}^{\alpha}_{\;\;\;\beta} \delta \Omega^{\prime\; (z,{\bar z}, \beta)} 
+h^{(z,{\bar z},\alpha )} \;, \nonumber
\end{eqnarray}
\begin{eqnarray}
\delta {\bar A}^{(z,{\bar z},\alpha)}=
-{\rm i} {\overline {\cal D}}^{\alpha}_{\;\;\;\beta} \delta \Phi^{(z,{\bar z}, \beta)}  
+{\overline {\cal D}}^{\alpha}_{\;\;\;\beta} \delta \Omega^{\prime\; (z,{\bar z}, \beta)}   
+{\bar h}^{(z,{\bar z},\alpha )}\;.
\label{better-holomorphic}
\end{eqnarray}
Notice that (\ref{absorb}) is
(\ref{2d-ym-redef}) and (\ref{better-holomorphic}) is (\ref{2d-ym-other-formula}) in holomorphic coordinates. It is 
easy to check that replacing (\ref{holomorphic}) by (\ref{better-holomorphic}) does not
affect (\ref{remarkable}) through (\ref{explicit-integral-formula}). This is because these
are gauge-invariant expressions, hence unaffected by the gauge transformation (\ref{absorb}).

\section{More about the (2+1)-dimensional Hamiltonian}
\setcounter{equation}{0}
\renewcommand{\theequation}{9.\arabic{equation}}

We now reexamine the form of the kinetic term of the Hamiltonian of a 
non-Abelian gauge theory in 2+1 dimensions, briefly 
discussed in Section 7, in the light of the result (\ref{remarkable}). Our
expressions are very similar those in references \cite{kar-nair,kar-kim-nair}.

The Yang-Mills Hamiltonian operator in two space dimensions is 
\begin{eqnarray}
H=T+U\;, \;\;\ T=-\frac{e^{2}}{2}\int \frac{\delta^{2}}{\delta A^{(x,\alpha)}\delta A^{(x,\alpha)}}
\;,\;\; U=\frac{1}{2e^{2}}\int (F_{12})^{2} \;, \label{holo-ym-hamiltonian}
\end{eqnarray}
and wave functionals depend only on $\Phi$. The coupling constant $e$ has
engineering dimension one-half. 

The kinetic term may be written with the aid of
(\ref{remarkable}) as the Laplacian 
\begin{eqnarray}
T
=-\frac{e^{2}}{2}\int \;
\frac{1}{\sqrt{G}}
\frac{\delta}{\delta\Phi^{(y,\beta)}}\; {\sqrt{G}}G^{(y,\beta)(z,\gamma)}\;
\frac{\delta}{\delta\Phi^{(z,\gamma)}}\;,
\label{holo-phi-ym-kinetic}
\end{eqnarray}
where the inverse metric tensor on the hypersurface is
\begin{eqnarray}
G^{(y,\beta)(z,\gamma)}=
\left[\frac{1}{\cal D}\frac{1}{\overline {\cal D}} \right]^{\beta \gamma} \delta^{2}(y-z)\;.
\label{holo-inv-metric tensor}
\end{eqnarray}
This tensor is therefore the inverse of (\ref{2d-metric-tensor}) after a change
of coordinates. Therefore, the metric tensor now reads
\begin{eqnarray}
G_{(y,\beta)(z,\gamma)}=
\left[{\cal D}{\overline {\cal D}} \right]_{\beta \gamma} \delta^{2}(y-z)\;.
\label{holo-metric tensor}
\end{eqnarray}
The inner product on Schr\"{o}dinger wave functionals is the functional
integral
\begin{eqnarray}
\left< \Psi_{1} \vert \Psi_{2} \right>
=\int [d\Phi]  \;{\sqrt {{\rm det}[{\cal D}{\overline {\cal D}}]}}\; {\overline {\Psi}}_{1}[\Phi] \;\Psi_{2}[\Phi]\;.
\label{holo-inner-product}
\end{eqnarray}
The kinetic term in equations (8.2a-e) of the first paper
of reference
\cite{kar-kim-nair} closely resembles (\ref{holo-phi-ym-kinetic}). It will 
be important to fully understand the connection.

\section{The dual Hamiltonian and the mass gap}
\setcounter{equation}{0}
\renewcommand{\theequation}{10.\arabic{equation}}

The origin of the Yang-Mills mass gap in the picture put forth in
references \cite{kar-nair,kar-kim-nair,nair-yel} is in the functional 
measure in the inner product. We shall briefly discuss this in the context of a dual
Hamiltonian formulation. We show only in principle how the gap can be 
obtained; the complete calculation is under study.

We first discuss quantum mechanics in curved space. Consider an 
n-dimensional manifold without boundary
with 
coordinates $x^{\mu}$, $\mu=1,\dots,n$ and metric tensor $g_{\mu \nu}$. The inner product is
\begin{eqnarray}
\left< \Psi_{1} \vert \Psi_{2} \right>= \int d^{n} x {\sqrt{{\rm det}g}}\; {\overline \Psi}_{1}(x) \Psi_{2}(x)
\;,\label{in-prod}
\end{eqnarray}
and the matrix element of minus the Laplace operator is
\begin{eqnarray}
\left< \Psi_{1} \vert -\Delta \vert \Psi_{2} \right>= 
\!&\!-\!&\!\int d^{n} x {\sqrt{{\rm det}g}}\; {\overline \Psi}_{1}(x) \frac{1}{\sqrt{{\rm det}g}}
\partial_{\mu}  {\sqrt{{\rm det}g}} g^{\mu \nu} \partial_{\nu} \Psi_{2}(x) \nonumber \\
\!&\!=\!&\!\int d^{n} x  {\sqrt{{\rm det}g}} \; g^{\mu \nu} \partial_{\mu} {\overline \Psi}_{1}(x) 
\partial_{\nu} \Psi_{2}(x) 
\;,\label{laplace-element}
\end{eqnarray}

Next we introduce 
$\psi_{1,2}(x)=({\rm det} g)^{\frac{1}{4}} \Psi_{1,2}(x)$, so that (\ref{in-prod}) has the standard flat-space form
\begin{eqnarray}
\left< \Psi_{1} \vert \Psi_{2} \right>= \int d^{n} x \;{\overline \psi}_{1}(x) \psi_{2}(x)
\;,\label{in-prod1}
\end{eqnarray}
and (\ref{laplace-element}) becomes
\begin{eqnarray}
\left< \Psi_{1} \vert -\Delta \vert \Psi_{2} \right> 
= \int d^{n} x \; g^{\mu \nu} \left[
\left( \partial_{\mu}-\frac{1}{4} \partial_{\mu}{\rm Tr}\log g \right)
{\overline \psi}_{1}(x) \right]
\left[ \left( \partial_{\nu}- \frac{1}{4} \partial_{\nu}{\rm Tr}\log g \right) 
\psi_{2}(x) \right] \nonumber
\end{eqnarray}
\begin{eqnarray}
=\int d^{n} x \;  \left\{g^{\mu \nu}  [\partial_{\mu}{\overline \psi}_{1}(x)]  
[\partial_{\nu} \psi_{2}(x)] 
+W(x){\overline \psi}_{1}(x) \psi_{2}(x) \right\}
\;,
\label{laplace-element1}
\end{eqnarray}
where in the last step we integrated by parts and
\begin{eqnarray}
W(x)
\!&\!=\!&\!
\frac{1}{4} \partial_{\mu}(g^{\mu \nu}\partial_{\nu} {\rm Tr}\log g)
+\frac{1}{16}g^{\mu \nu} (\partial_{\mu}{\rm Tr}\log g)(\partial_{\nu} {\rm Tr}\log g) \nonumber \\
\!&\!=\!&\!
({\rm det} g)^{-\frac{1}{4}}\partial_{\mu} [g^{\mu \nu}\partial_{\nu}
({\rm det} g)^{\frac{1}{4}}]\;.
\label{w-definition}
\end{eqnarray}

Consider now a quantum-mechanical 
Hamiltonian $H=-\Delta +U(x)$. In
the neighborhood
of a point of the manifold 
where $g_{\mu \nu}=\delta_{\mu \nu}$, we can expand $g^{\mu \nu}=\delta^{\mu \nu}+\cdots$,
the function $W(x)$ can be interpreted as a correction to the potential energy $U(x)$. This is
analogous to
what is done in references \cite{kar-nair, kar-kim-nair, nair-yel} in their coordinates
and what we
shall do in our own coordinates. The only difference is that the manifold has an infinite
number of dimensions. We define the effective Hamiltonian in finite dimensions as the operator with
matrix elements
\begin{eqnarray}
\left<\psi_{1} \vert H_{\rm eff} \vert \psi_{2} \right>
=\int d^{n} x \; \left\{\frac{1}{2} g^{\mu \nu}  [\partial_{\mu}{\overline \psi}_{1}(x)]  
[\partial_{\nu} \psi_{2}(x)] 
+[\frac{1}{2}W(x)+U(x)]{\overline \psi}_{1}(x) \psi_{2}(x) \right\} \;, \label{effective-hamiltonian}
\end{eqnarray}
where $\psi_{1}$ and $\psi_{2}$ are the square-integrable ``wave functions" with
inner product (\ref{in-prod1}).

Next we write down the analogue of (\ref{effective-hamiltonian}) for Yang-Mills theory in 2+1 dimensions. We
write the metric (\ref{2d-metric-tensor}) as
\begin{eqnarray}
G_{(x,\alpha)(y,\beta)}= -\partial^{2}\delta_{\alpha \beta}\delta^{2}(x-y)
+\left[(-{\cal D}^{2}+\partial^{2}{\rm 1}\!{\rm I} ) +{\cal F}_{12}\frac{1}{-{\cal D}^{2}}
{\cal F}_{12}\right]_{\alpha \beta}\delta^{2}(x-y) \;, \nonumber
\end{eqnarray}
and its inverse as an expansion
\begin{eqnarray}
G^{(x,\alpha)(y,\beta)}= \frac{1}{-\partial^{2}} \delta^{\alpha \beta}\delta^{2}(x-y)
\;\;\;\;\;\;\;\;\;\;\;\;\;\;\;\;\;\;\;\;\;\;\;\;\;\;\;\;\;\;\;\;\;\;\;\;\;\;\;\;\;\;
\;\;\;\;\;\;\;\;\;\;\;\; \nonumber
\end{eqnarray}
\begin{eqnarray}
\;\;\;\;\;\;\;\;\;\;\;\;
- \left( \frac{1}{-\partial^{2}} \right)
\left[  (-{\cal D}^{2}+\partial^{2} {\rm 1}\!{\rm I} )+{\cal F}_{12}\frac{1}{-{\cal D}^{2}}
{\cal F}_{12} \right]_{\alpha \beta} \left( \frac{1}{-\partial^{2}} \right)
\delta^{2}(x-y)+\cdots \;, \label{2d-expansion-inverse-metric}
\end{eqnarray}
where $({\rm 1}\!{\rm I})_{\alpha \beta}=\delta_{\alpha \beta}$. We 
introduce a new physical field ${\cal Y}^{\alpha} = ( -\partial^{2} )^{1/2} \Phi^{\alpha}$. We assume
that $\Phi^{\alpha} = \delta \Phi^{\alpha}$ is close to zero and identify zero gauge
field with zero $\Phi$. In this approximation the magnetic
field is
\begin{eqnarray}
{\cal F}_{12}\!&\!=\!&\!
\delta {\cal F}_{12}= 
{\cal D}_{1} \delta A_{2} - {\cal D}_{2} \delta A_{1}
=\left( -{\cal D}^{2}- {\cal F}_{12} \frac{1}{-{\cal D}^{2}}  {\cal F}_{12}
\right)  \delta \Phi  \nonumber \\
\!&\!=\!&\!
\left( -{\cal D}^{2}- {\cal F}_{12} \frac{1}{-{\cal D}^{2}}  {\cal F}_{12}
\right) ( -\partial^{2} )^{-1/2}   {\cal Y}  =( -\partial^{2} )^{1/2}   {\cal Y} +\cdots
 \;. \label{2d-field-strength-var}
\end{eqnarray}
The effective Hamiltonian takes the form
\begin{eqnarray}
H_{\rm eff}=\int \left\{ -\frac{e^{2}}{2}\frac{\delta^{2}}{ (\delta {\cal Y}^{\alpha})^{2}}
+\frac{e^{2}}{2} W[{\cal Y}] +\frac{1}{2e^{2}} (\partial_{j} {\cal Y}^{\alpha})^{2}+\cdots  \right\}\;,
\label{2d-ym-effective}
\end{eqnarray}
where the higher-order terms can be found from (\ref{2d-expansion-inverse-metric}) and (\ref{2d-field-strength-var})
and
\begin{eqnarray}
W[ {\cal Y} ]=
({\rm det} G)^{-\frac{1}{4}}  \frac{\delta}{ \delta \Phi^{(x,\alpha)} } 
\left[   G^{(x,\alpha)(y,\beta)} \frac{\delta}{\delta \Phi^{(y,\beta)}}  ({\rm det} G)^{\frac{1}{4}}
\right] \;. \label{2d-big-w-definition}
\end{eqnarray}
The leading terms of the effective Hamiltonian (\ref{2d-ym-effective}) resemble an ordinary scalar field theory. The
gap is a term of the form $({\cal Y}^{\alpha})^{2}$ in $W[{\cal Y}]$. Its existence depends on the form
of the regularized determinant of the metric in (\ref{2d-big-w-definition}).

An expression for the effective Hamiltonian can also
be obtained in 3+1 dimensions. The leading terms are those of an Abelian gauge theory, instead
of a scalar field theory. The appearance of the gap would be a signal of the breaking of dual gauge symmetry.

We write the three-dimensional metric tensor as
\begin{eqnarray}
G_{(x,l,\alpha)(y,m,\beta)} = 
(-\partial^{2}\delta_{lm}-\partial_{l}\partial_{m})\delta_{\alpha \beta} \delta^{3}(x-y)
\;\;\;\;\;\;\;\;\;\; 
\;\;\;\;\;\;\;\;\;\; 
\;\;\;\;\;\;\;\;\;\; 
\;\;\;\;\;\;\;\;\;\; 
\;\;\;\;\;\;\;\;\;\; 
\nonumber 
\end{eqnarray}
\begin{eqnarray}
\;\;\;\;\;\;+
\left[(-{\cal D}^{2}+\partial^{2}{\rm 1}\!{\rm I} \delta^{lm}) +
({\cal D}_{l}{\cal D}_{m}-\partial_{l}\partial_{m} {\rm 1}\!{\rm I} )
-{\rm i} {\cal F}_{lm}
+ (\ast {\cal F})_{l}\frac{1}{-{\cal D}^{2}}(\ast {\cal F})_{m}
\right]_{\alpha \beta}\delta^{3}(x-y)\;. \nonumber
\label{3d-metric-expansion}
\end{eqnarray}
There are several ways to dual-gauge fix, thereby removing the zero mode of the first term of
the right-hand side of (\ref{3d-metric-expansion}). For example, taking a dual-Coulomb gauge 
$\partial_{l}\Phi^{(x,l,\alpha)}=0$ will lead to the expansion
of the inverse metric tensor beginning as
\begin{eqnarray}
G^{(x,l,\alpha)(y,m,\beta)} = \left(\frac{1}{-\partial^{2}}\right)\delta^{\alpha \beta} \delta^{3}(x-y)
+\cdots \;.
\label{3d-inv-metric-exp}
\end{eqnarray}
The form of this expression indicates that we should introduce the physical field ${\cal Y}^{(x,l,\alpha)}
=(-\partial^{2})^{1/2}\Phi^{(x,l,\alpha})$.

The field strength for small $\{\Phi\}$ may be approximated as
\begin{eqnarray}
{\cal F}_{jk}^{\alpha} \approx \delta {\cal F}_{jk}^{\alpha}
=\left[
\epsilon^{kml}{\cal D}_{j}{\cal D}_{m} -\epsilon^{jml}{\cal D}_{k}{\cal D}_{m}
+{\cal F}_{jk}\frac{1}{{\cal D}^{2}} (\ast {\cal F})_{l}
\right]_{\alpha \beta} \delta \Phi^{(x,l,\beta)} \;.
\nonumber
\end{eqnarray}
The expansion of the potential energy in dual-Coulomb gauge is therefore
\begin{eqnarray}
\int \frac{1}{4}\left( {\cal F}_{jk}^{\alpha} \right)^{2} = \int \frac{1}{2}
\Phi^{(x,l,\alpha)} (\partial^{2})^{2} \Phi^{(x,l,\alpha)}
=\int \frac{1}{2} {\cal Y}^{(x,l,\alpha)} (-\partial^{2}) {\cal Y}^{(x,l,\alpha)}\;.
\label{3d-pe}
\end{eqnarray}
The effective Hamiltonian in 3+1 dimensions is
\begin{eqnarray}
H_{\rm eff}=\int \left\{ -\frac{e_{0}^{2}}{2}\frac{\delta^{2}}{ (\delta {\cal Y}^{(x,l,\alpha)})^{2}}
+\frac{e^{2}}{2} {\rm W}[{\cal Y}] +\frac{1}{2e_{0}^{2}} (\partial_{j} {\cal Y}^{(x,l,\alpha)})^{2}+\cdots  \right\}\;,
\label{3d-ym-effective}
\end{eqnarray}
where W is now
\begin{eqnarray}
{\rm W}[ {\cal Y} ]=
({\rm det} G)^{-\frac{1}{4}}  \frac{\delta}{ \delta \Phi^{(x,l,\alpha)} } 
\left[   G^{(x,l,\alpha)(y,m,\beta)} \frac{\delta}{\delta \Phi^{(y,m,\beta)}}  ({\rm det} G)^{\frac{1}{4}}
\right] \;.  \label{3d-big-w-definition}
\end{eqnarray}
The mass gap is obtained by expanding W, as in 2+1 dimensions. 

Notice that the form of the effective Hamiltonian (\ref{3d-ym-effective}) will change if we impose a
a different gauge condition on the dual gauge field. Furthermore
the right-hand side of (\ref{3d-big-w-definition}) is clearly dual-gauge dependent. This is how the dual
gauge invariance is preserved. 

\section{Conclusions and Outlook}

Through a non-Abelian decomposition formula for differential one-forms, we have found a 
special hypersurface in the space of connections. Gauge transformations are normal
to this hypersurface, which makes it a useful tool in the geometry
of gauge configurations. We have found expressions for
the curvature tensor and sectional, Ricci and scalar curvatures of
this hypersurface for (D+1)-dimensional gauge theories. The
hypersurface coordinate was explicitly determined 
in terms of the gauge connection in 2+1 dimensions. A Hamiltonian formalism 
close to that of Karabali, Kim and Nair was worked out. Finally, we discussed
an effective Hamiltonian of the dual degrees of freedom.

There are many avenues which remain
to be explored. Though our methods are
evidently related to Karabali, Kim and Nair's (2+1)-dimensional formalism,
the connection is not complete. We do not yet have a mapping between their degrees
of freedom and ours. Furthermore, we do not yet see the connection between our 
ideas and Nair and Yelnikov's \cite{nair-yel} in 3+1 dimensions. A mass for the 
dual gauge field in 3+1 dimensions implies 
a type of dual Higgs phase. Though our duality is not 
the same as Kramers-Wannier duality, we believe that 
this would imply that the vacuum is a chromoelectric
superconductor \cite{mand-thooft}. We hope that this issue will become clearer as
we understand dual gauge invariance more completely.

The most urgent project is to regularize the determinant of the metric (in 2+1 and
3+1 dimensions) in order to 
obtain the dual Hamiltonian more explicitly. In this way, the mass gap and perhaps
much of the physics can be understood, at least at strong coupling.

The geometry of the hypersurface needs to be examined
more closely. We have discussed how to compute the Ricci tensor and curvature scalar in dimensional
regularization, but the details of this computation are not finished.

Though we have mentioned the use of dimensional
regularization in Section 6, we 
have not extensively discussed cut-off methods in this paper. A simple gauge-invariant
regulator is to drop high-momentum (short-wavelength) components from $\{\Phi\}$. There is
no obvious inconsistency with this idea in 2+1 dimensions. In 3+1 dimensions, such a procedure
may explicitly break dual gauge invariance; perhaps this is not a serious
difficulty, and it is enough to have ordinary gauge invariance. We 
have already explored a lattice version of our method, where all gauge invariances are preserved. In a
Hamiltonian lattice formalism, where the gauge group is
enlarged to GL($N,{\rm C}\!\!\!{\rm I}$), then broken to
SU($N$), a linear transformation $L$ can be defined. Another lattice formulation,
with approximate conformal invariance of the kinetic term in 2+1 dimensions, has been presented in
a new paper by
Rajeev \cite{rajeev}. It should be quite interesting to see whether an analogue of $L$ exists in
his approach.

\section*{Acknowledgements}

We thank both D. Karabali and V.P. Nair for discussions concerning the relation between the 
work presented here and
that in references \cite{kar-nair,kar-kim-nair, nair-yel} and for comments on the manuscript.


\begin{thebibliography}{xx}
\bibitem{foot1} In two space dimensions, $L$ 
maps scalar Lie-algebra valued
functions to vector Lie-algebra valued functions. Hence
it has, strictly speaking, no self-adjoint
definition. However, fixing $j$ in
${\rm i}L^{(x,j,\alpha)}_{\;\;\;\;\;\;(y,\beta)}$ defines operators
which should have self-adjoint extensions. The same
idea should work for other numbers of space dimensions.
\bibitem{foot2} Formally such solutions can always be obtained by
integrating these equations for $b$ and $\bar b$. These solutions will not be normalizable
and single-valued, in general, hence they are irrelevant.
\end{thebibliography}
\end{document}